\documentclass[a4paper]{IEEEtran} %alt waldir (para acrescentar o heading)

\ifCLASSOPTIONcompsoc
  \usepackage[nocompress]{cite}
\else
  \usepackage{cite}
\fi
\usepackage{scalefnt}
\usepackage{amssymb}
\usepackage{amsthm}
\usepackage{xcolor}
\newtheorem{mydef}{Definition}
\usepackage[linesnumbered,lined,boxed,commentsnumbered]{algorithm2e}
\ifCLASSINFOpdf
  \usepackage[pdftex]{graphicx}
  \graphicspath{{../pdf/}{../jpeg/}}
   \DeclareGraphicsExtensions{.pdf,.jpeg,.png}
\else
   \usepackage[dvips]{graphicx}
   \graphicspath{{images/}}
   \DeclareGraphicsExtensions{.eps}
\fi
\usepackage{amsmath}
\def\baselinestretch{0.96}

\begin{document}
\title{Verification of Magnitude and Phase Responses \\ in Fixed-Point Digital Filters}

\author{\IEEEauthorblockN{Daniel P. M. de Mello$^1$, Mauro L. de Freitas$^2$, Lucas C. Cordeiro$^1$, Waldir S. S. J\'{u}nior$^1$,\\ Iury V. de Bessa$^1$, Eddie B. L. Filho$^{1,3}$ and Laurent Clavier$^2$ \\}
\IEEEauthorblockA{$^1$Universidade Federal do Amazonas (UFAM), Manaus-AM, Brasil \\
$^2$Univ. Lille, CNRS, ISEN, Univ. Valenciennes, UMR 8520, IEMN,
         59000 Lille, France \\
$^3$Samsung Instituto de P\&D da Amaz\^onia (SIDIA), Manaus-AM, Brasil \\
Email: dani-dmello@hotmail.com, m.lopesdefreitas@ed.univ-lille1.fr, 
 lucascordeiro@ufam.edu.br, \\ waldirjr@ufam.edu.br, iurybessa@ufam.edu.br, eddie.l@samsung.com, laurent.clavier@iemn.univ-lille1.fr}
}


% compsoc conferences compared to traditional conferences), use this
%Telephone: (800) 555--1212, Fax: (888) 555--1212}


\pagenumbering{gobble} %alt waldir (retirar num de pagina)

\maketitle

\markboth{XXXV SIMP\'OSIO BRASILEIRO DE TELECOMUNICA\c{C}\~OES E PROCESSAMENTO DE SINAIS - SBrT2017, 3-6 DE SETEMBRO DE 2017, S\~AO PEDRO, SP} {XXXV SIMP\'OSIO BRASILEIRO DE TELECOMUNICA\c{C}'\OES E PROCESSAMENTO DE SINAIS - SBrT2017, 3-6 DE SETEMBRO DE 2017, S\~AO PEDRO, SP}

\begin{abstract}
In the digital signal processing (DSP) area, one of the most important tasks is digital filter design. Currently, this procedure is
performed with the aid of computational tools, which generally assume filter coefficients represented with floating-point arithmetic.
Nonetheless, during the implementation phase, which is often done in digital signal processors or field programmable gate arrays, the representation of
the obtained coefficients can be carried out through integer or fixed-point arithmetic, which often results in unexpected behavior or even unstable filters.
The present work addresses this issue and proposes a verification methodology based on the digital-system verifier (DSVerifier), with the goal of checking fixed-point digital filters w.r.t. implementation aspects. In particular, DSVerifier checks whether the number of bits used in coefficient representation will result in a filter with the same features specified during the design phase. Experimental results show that errors regarding frequency response and overflow are likely to be identified with the proposed methodology, which thus improves overall system's reliability. 
\end{abstract}


%  \ifCLASSOPTIONpeerreview
%  \begin{center} \bfseries EDICS Category: 3-BBND \end{center}
%  \fi
\IEEEpeerreviewmaketitle

\section{Introduction}
Digital Filters with finite impulse response (FIR) 
or infinite impulse response (IIR) are used in different 
areas, such as digital signal processing (DSP), control 
systems, telecommunications, medical instrumentation, and 
consumer electronics. In general, such applications vary 
from simple frequency selection and adaptive 
filters to equalizers and filter banks, whose objective 
is to modify the characteristics of a certain signal, in 
accordance with pre-established requisites. 

Digital filter design follows an abundant 
mathematical theory, both in frequency and time 
domains, and is usually realized with tools as MATLAB~\cite{matlab}, 
which normally assume fixed- or floating-point precision. 
Nonetheless, there can be a great disparity between a 
filter design and its practical 
implementation. For instance, many projects are 
implemented in digital signal processors (DSPS) 
or field programmable gate arrays (FPGA), which can employ 
finite precision, based on fixed-point arithmetic 
(with lower cost and complexity), whereas the associated design 
usually assume floating-point precision.

This difference has the potential of generating 
undesirable effects regarding a filter's frequency response, 
both in phase and magnitude, in addition to problems as overflow and 
instability. Such behavior is due to quantization errors caused by finite precision, 
which result in coefficients that are different from the ones
originally designed. As a result, there might be 
questions, in the implementation phase, regarding the 
effectiveness of digital filters and the number of bits 
needed for their representation, in such a way that design 
parameters are also satisfied.  

This article presents a verification methodology for 
digital filters with fixed-point implementation, based on the
Efficient SMT-Based Context-Bounded Model Checker (ESBMC), 
which employs Bounded Model Checking (BMC) techniques and 
Satisfiability Modulo Theories  (SMT)~\cite{smtbased,Cordeiro2011}. 
Such an approach indicates, according to previously defined filter parameters, if the chosen number of bits is sufficient and does not lead to unexpected errors 
or behaviors. The main advantage of this approach, over 
other filter analysis techniques~\cite{Oppen1999,Padgett}, is that 
model-checking tools can provide precise information 
on how to reproduce errors (for instance, system input 
values) through counterexamples.

In order to apply the proposed methodology to digital filter verification, 
the Digital System Verifier (DSVerifier) tool was used, which is a 
front-end tool for the verification of different types of 
digital systems, with the aid of BMC techniques. 
An implementation written in C was developed and integrated 
into DSVerifier~\cite{Abreu2016,Ismail2015}, because it was unable to support the verification of 
filter-specific properties, such as magnitude and phase, prior to this work. 
Various kinds of practical digital filters were 
used for verification, with the goal of validating them against real designs.
As a result, such a verifier, together with traditional design tools, 
provide a complete digital filter synthesizing scheme, according to application conditions.

Indeed, the present approach is effective in verifying magnitude and phase responses, 
which provides an analysis deeply based on DSP theory and 
fills gaps presented in the existing literature. The performed experiments are based on a set of publicly available 
benchmarks.\footnote[2]{The benchmarks are available at http://www.esbmc.org/benchmarks/sbrt2017.zip}

This work is organized as follows. Section II presents 
the verification schemes available in the literature, 
highlighting its main characteristics. In section III, the 
BMC technique is presented. Then, in section IV, the 
proposed method is described, and section V presents the 
simulations results. Finally, the conclusions are set out 
in section VI.

%\textit{Outline}. Section II presents 
%the verification schemes available in the literature, 
%highlighting its main characteristics. In Section III, the 
%BMC technique is presented. Then, in Section IV, the 
%proposed method is described, while Section V presents the 
%experimental results. Finally, conclusions and future work
%are described in Section VI.

%------------------------------
\section{Related Work}
%------------------------------

The application of tools that implement the BMC technique, 
regarding software verification, is becoming quite popular, 
mainly due to the advent of sophisticated SMT solvers, which are
constructed based on efficient satisfiability solvers (SAT)~\cite{z3}. 
Previously published studies related to SMT-based BMC, for software, 
handle the problem of verifying ANSI-C programs that use bit 
operations, fixed- and floating-point arithmetic, comparisons 
and pointers arithmetic~\cite{smtbased}; however, there is 
little evidence of studies that address the verification of properties 
related to digital filters implementation, in ANSI-C, especially when 
assuming arbitrary word-length. One of such studies was previously conducted 
by Freitas {\it et al.} ~\cite{sbrt13}, where digital filter properties, 
such as overflow, magnitude and stability, were verified by employing ESBMC. 
Those results served as main inspiration for the present work, whose proposal is to further 
extend and reproduce the verified properties, while creating a support for them on the DSVerifier tool. Now this new implementation allows passband filters verifications. 

Akbarpour and Tahar~\cite{Akbarpour2004,Akbarpour2007} presented a mechanical approach for error detection in digital-filter design, which is based on a high-order logic (HOL) theorems solver. The authors describe valuation functions that find the real values of a digital filter's output, through fixed- and floating-point representations, aiming to define an error. The latter represents the difference between the encountered values, through this valuation function, and the output corresponding to design specifications.

Recently, Cox, Sankaranarayanan and Chang~\cite{Cox2012}
introduced a new approach that uses precise bit 
analysis for the verification of digital filter 
implementations, in fixed-point. This approach 
is based on the BMC technique and employs SMT 
solvers for checking verification conditions, 
which are generated in the digital-filter design phase. 
The authors show that such an approach is more 
efficient and produces fewer false alarms, if 
compared to those that use real arithmetic solvers; however, the mentioned studies do not address 
intrinsic filters characteristics, 
such as errors or modifications related to 
poles, zeros, or frequency response.

Abreu {\it et al.} also proposed a new methodology for the
verification of digital filters, named as DSVerifier, 
which is based on state-of-the-art bounded model checkers that support full 
C and employ solvers for boolean satisfiability and satisfiability modulo theories~\cite{Abreu2016}. 
In addition to verifying overflow and limit-cycle occurrences, DSVerifier can 
also check output errors and time constraints, based on discrete-time models implemented in C.

The findings presented in the previous 
studies served as inspiration for the 
present article, which aims to extend the 
approach proposed by Cox, Sankaranarayanan 
and Chang~\cite{Cox2012} and Abreu {\it et al.}~\cite{Abreu2016} includes new digital filter properties to be checked, such 
as magnitude and phase responses, which provides an analysis closed related to DSP theory.  In addition, this work applies DSVerifier, which is a model checking tool to investigate finite word-length (FWL) effects in digital systems implementations, to the verification of a more diverse set of benchmarks, including different classes of filters ({\it e.g.}, passband filters).

%---------------------------------
\section{The BMC technique}
%---------------------------------

With ESBMC, a program under analysis is modeled 
by a state transition system, which is generated 
from the program control-flow graph (CFG)~\cite{CFG} that is automatically 
created during the verification process. 
A node in a CFG represents an assignment 
(deterministic or nondeterministic) or a conditional 
expression, while an edge represents a change in a program's flow. 

A state transition system $M = (S, T, S_0)$
is an abstract machine, which consists in a state set
$S$, where $S_0 \subseteq S$ represents a initial state
set and $T \subseteq S \times S$ is the transition relation. 
A state $s \in S$ consists of the value of a program counter 
$pc$ and also the values of all variables in an application. %DÚVIDA DE TRADUÇÃO. ORIGINAL: Aplicativo. Tradução: Aplication?%
An initial state $s_0$ assigns the program's initial location $\gamma = (s_i, s_{i+1})\in T$, in the CFG.
The transitions are identified as 
between two states $s_i$ and $s_{i+1}$, with a logical formula 
$\gamma (s_i , s_{i+1} )$ that contains the value restrictions of the 
program counter and the system's variables.

Given a transition system $M$, a property $\phi$ and a bound $k$, ESBMC unfolds 
a system $x$ times and transforms the associated result into a verification condition $\psi$, in
such a way that $\psi$ is satisfiable if $\phi$ contains a counterexample
with length smaller than $x$~\cite{smtbased}. Thus, the BMC technique problem is formulated as follows

\begin{equation}
\footnotesize
\label{bmceq}
\psi_x = I(s_0) \wedge \bigvee^x_{i=o} \bigwedge^{i-1}_{j=0} \gamma(s_j, s_{j+1}) \wedge ¬\phi(s_i),
\end{equation}

\noindent where $\phi$ is a property, $I$ is a set of the initial states in $M$, and $\gamma (s_j , s_{j+1} )$ 
is the state transition function of $M$ between steps $j$ and $j+1$. 
Thus, $I(s_0) \wedge \bigwedge^{i-1}_{j=0} \gamma(s_j, s_{j+1})$ represents the execution
of $M$ for $i$ times and eq. \eqref{bmceq} will only be satisfied if, and only if, for each $i \leq x$, there is a 
reachable state where $\phi$ is violated. If eq. \eqref{bmceq} is satisfiable, then ESBMC shows a counterexample,
 defining which variable values are needed to lead to the related error. The counterexample for a property 
 $\phi$ is a state sequence $s_0, s_1, ..., s_x $ with $ s_0 \in S_0$ and $\gamma (s_i, s_{i+1})$ , for
$0 \leq i < x$. If eq. \eqref{bmceq} is not satisfied, one can conclude that no error stateis reachable with $x$ steps or less. 
 
%------------------------------------------------- 
\section{The New Verification Methodology}
%-------------------------------------------------

In general, the fixed-point implementation uses standard 
registers to store the inputs and outputs along the adders, multipliers,
and delays. However, the results of these elements might exceed the 
limits of the allocated variables, or generate different values than expected,
due to the coefficients accuracy or the associated number of bits.
As a result, it is possible that the result differs from the one specified 
in the design or even that a filter becomes unstable without this occurring 
in the filter design.

With this in mind, the proposed verification methodology is split 
into three main parts: magnitude and phase verification, poles and zeros stability, 
and overflow verification.

%--------------------------------------------------
\subsection{Magnitude and phase verification}
%-------------------------------------------------- 

\begin{figure}[ht]
    \centering
    \includegraphics[width=0.8\columnwidth]{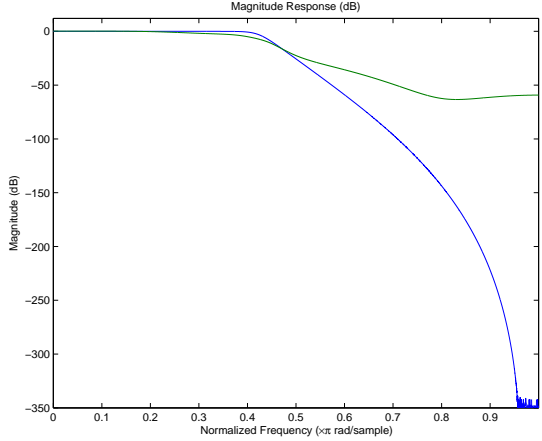}
    \caption{Magnitude response of IIR Chebyshev filter with order 12. The blue curve represents the projected answer and the green curve the response with fixed-point containing one sign bit, $7$ bits for the integer and $6$ for the fractional parts} \vspace*{-2ex}
    \label{fig:iir_lp12-fixed-8-6}
\end{figure}

Changes in the coefficients, due to the fixed-point quantization, 
alter the response in magnitude and phase~\cite{Akbarpour2007}. 
An example of this can be seen in Fig.~\ref{fig:iir_lp12-fixed-8-6}.

In this first approach, the input of the proposed verification system is composed by the filter coefficients in floating-point by the design properties,
which must be analyzed according to the adopted conditions, such as passband, cut-off frequency, rejection band, as well as the gains in each region and the amount of bits used for the representation in the fixed-point.

Given that $N$ is the number of points of the Discrete-Time Fourier Transform (DTFT)~\cite{Diniz2002}, $h[n]$ is the filter impulse response and $H_k$ is the k-th component of it's sampled equivalent in frequency domain, we have that
\begin{equation}
\footnotesize
H_k = \sum\limits_{n=0}^{N-1} h(n)e^{-j(2\pi/N)kn}.
\end{equation}

In addition, suppose that $\omega_{p}$, $\omega_{r}$ and $\omega_{c}$ are the digital frequencies of passband, stopband and cutoff, respectively. In turn, $A_{p}$, $A_{r}$ and $A_{c}$ are the gains that will be checked. We assumed the following assertions to verify magnitude and phase properties for lowpass and highpass filters.
{\footnotesize
\begin{align}
 I_{lp\_mag} \Leftrightarrow &\bigg[ \Big(\arrowvert H_k \arrowvert < A_{p} \Big) \land \Big(0 \le \tfrac{2 \pi k }{N} \le \omega_{p} \Big) \bigg] \ \vee \nonumber \\
&\bigg[ \Big(\arrowvert H_k \arrowvert > A_{c} \Big) \land \Big(\tfrac{2 \pi k }{N} = \omega_{c} \Big) \bigg] \ \vee \nonumber \\
&\bigg[ \Big(\arrowvert H_k \arrowvert > A_{r} \Big) \land \Big(\omega_{r} \le \tfrac{2 \pi k }{N} \le \pi \Big) \bigg],
\end{align}
}%
{\footnotesize
\begin{flalign}
 I_{hp\_mag} \Leftrightarrow &\bigg[ \Big(\arrowvert H_k \arrowvert > A_{r} \Big) \land ~ \Big(0 \le \tfrac{2 \pi k }{N} \le \omega_{r} \Big)\bigg] \ \vee \nonumber \\ 
&\bigg[ \Big(\arrowvert H_k \arrowvert < A_{c} \Big) \land \Big( \tfrac{2 \pi k }{N} = \omega_{c} \Big) \bigg] \ \vee \nonumber \\ 
&\bigg[\Big(\arrowvert H_k \arrowvert < A_{p} \Big) \land \Big(\omega_{p} \le \tfrac{2 \pi k }{N} \le \pi\Big) \bigg]
\end{flalign}
}%
\noindent and
% \begin{align}
%  I_{lp\_phase}, I_{hp\_phase} & \Leftrightarrow \arrowvert \sphericalangle H(k) - \sphericalangle H_{fixed}(k) \arrowvert > threshold.
% \end{align}
{\footnotesize
\begin{eqnarray}
  \begin{array}{ll}
       & I_{lp\_phase}, \\
       & I_{hp\_phase}
  \end{array}
   \Leftrightarrow \arrowvert \sphericalangle H(k) - \sphericalangle H_{fixed}(k) \arrowvert > threshold.
\end{eqnarray}
}%
\noindent
In case of assertion violation, an error is generated to indicate that the
amount of bits is insufficient for the representation, taking into
account the initial design restrictions.

%---------------------------------------------
\subsection{Poles and zeros verification} 
\label{polos-zeros}
%---------------------------------------------

Jury's algorithm is used to check the stability in the $z$-domain for a given characteristic polynomial of the form
\begin{equation}
\footnotesize
S(z) = a_{0}z^{N} + a_{1}z^{N-1} + ... a_{N-1}z + a_{N} = 0, a_{0}\neq 0
\end{equation}
In particular, Jury stability test is already explained in the control system literature\cite{IEEE17_tc}. This study, however, limits itself to explain the SMT encoding of Jury's criteria. For the stability test procedure, the following Jury matrix $M=[m_{ij}]_{(2N-2)\times N}$ is built from $S(z)$ coefficients:
\begin{equation}
\footnotesize
M=\left[
\begin{matrix}
  V^{(0)} \\
  V^{(1)} \\
  \vdots \\
  V^{(N-2)}
 \end{matrix}
\right]\mbox{,}
\end{equation}
\noindent where $V^{(k)}=[v^{(k)}_{ij}]_{2\times N}$ such that
\begin{equation}
\footnotesize
v^{(0)}_{ij}=\begin{cases} 
a_{j-1}, & \mbox{if } i=1 \\   v^{(0)}_{(1)(N-j+1)}, & \mbox{if } i=2 
\end{cases}\mbox{, and} 
\end{equation}
\begin{equation}
\footnotesize
v^{(k)}_{ij}=\begin{cases} 
0, & \mbox{if } j>n-k\\
v^{(k-1)}_{1j}-v^{(k-1)}_{2j}\cdot\frac{v^{(k-1)}_{11}}{v^{(k-1)}_{21}}, & \mbox{if } j\leq n-k  \mbox{ and } i=1 \\
v^{(k)}_{(1)(N-j+1)}, & \mbox{if } j\leq n-k \mbox{ and } i=2 \\
\end{cases} \mbox{,}
\end{equation}

\noindent where $k\in\mathbb{Z}$, such that $0<k<N-2$. $S(z)$ is the characteristic polynomial of a stable system if and only if the following four propositions hold:

\begin{itemize}
\footnotesize
\item $R_{1}$: $S(1)>0$;
\item $R_{2}$: $(-1)^{N}S(-1)>0$;
\item $R_{3}$: $\vert{a_{0}}\vert <a_{N}$;
\item $R_{4}$: $m_{11}>0\iff m_{31}\wedge m_{51}\wedge \dots \wedge m_{(2N-3)(1)}$.
\end{itemize}

The stability property is then encoded by creating a constraint using the fixed size bit-vector theory, typically supported by state-of-the-art SMT solvers~\cite{BarrettSST09}:

\begin{equation}
\footnotesize
\label{eq:stab_lit}
\phi_{stability}\iff \ (R_{1} \wedge R_{2} \wedge R_{3} \wedge R_{4}),
\end{equation}
\noindent where the literal $\phi_{stability}$ represents the validity of the stability condition; in particular, the SMT-solver checks whether Jury criteria hold for the characteristic polynomial coefficients.

%-------------------------------------------------------
\subsection{Overflow verification} \label{overflow}
%-------------------------------------------------------

The third part concerns the overflow verification after the coefficients quantization, 
which would be considered infeasible without computational tools.

The addition, subtraction, multiplication, and division operations allow the fixed-point 
representation; however, losing precision to respect the bits limitation amount. 
The overflow happens when the bits representation is violated.
In order to better understand it we will study two overflow types. 
{\small
\begin{mydef}
\label{saturationdef}
The saturation occurs when values outside of the bit range are represented by the minimum or maximum values.
\end{mydef}
}%
Despite the easy possibility to find the limits, it would be difficult to know which input will create the saturations.
{\small
\begin{mydef}
\label{wraparounddef}
The wrapping around occurs when the maximum value is attributed instead of a minimum value and vice-versa~\cite{Cox2012}.
\end{mydef}
}%
The input verification has nondeterministic fixed-point numbers $\tilde{x}[n]$,
the $h[n]$ filter coefficients that will be verified and the number of inputs $N$. All the overflow check iterations are given by
{\footnotesize
\begin{flalign}
I_{overflow} \Leftrightarrow \bigg[&\Big(\tilde{x}[n]h[0] > V_{max}\Big)~\vee \nonumber \\
                             & \Big(\tilde{x}[n-1]h[1] > V_{max}\Big)~\vee \nonumber \\
                             & \ldots \nonumber \\ 
                             & \Big(\tilde{x}[n-N-1]h[N-1] > V_{max}\Big)\bigg]
\end{flalign}
}%

To detect an error, a counterexample is generated. This counterexample consists 
of the violated states, providing access to the inputs which generated the error, 
in a specific order, as well as the output value. This approach gives to the filter 
designer information to understand the error conditions of overflow and underflow, 
allowing the desginer to come up with an alternative implementation.

%-------------------------------------
\section{Experiments} 
\label{simul}
%-------------------------------------

This section consists of two parts. The system configuration 
is described in Section \ref{exp1}, while Section \ref{exp2} summarizes our objectives with the experiments conducted and Section \ref{exp3} describes the results obtained with the DSVerifier tool\footnote[1]{http://www.dsverifier.org}~\cite{smtbased,Cordeiro2011,Ismail2015,Abreu2016}, implementing the changes regarding the filter magnitude and phase, as well as the preexisting functions for verification of stability and overflow.

%-------------------------------------
\subsection{System configuration and preparation for the experiments} 
\label{exp1}
%-------------------------------------

The set of magnitude and phase verification where split into two main groups: 
one consisting of IIR filters and another of FIR filters.
Each set is divided into 3 categories: lowpass, highpass and bandpass, 
with three filters of small order (2nd or 4th order), and three filters 
with high order (12th or 30th order) in each set for different cut-off frequencies. 
Three types of IIR filters were used: Butterworth, Chebyshev, and Elliptic. 
For FIR filters, the types Equiripple, Hann Window, and Maximally Flat. 

Altogether, $54$ stable filters were created during the design stage, 
with $18$ FIR and $36$ IIR, with sample frequency of $48kHz$. 
All filters transfer functions were obtained with the Filter Design and Analysis Tool 
application available in MATLAB~\cite{matlab} and written in a ``.c'' file according to the input specifications for DSVerifier. 

Aiming to explore different theories employed on the SMT~\cite{smtbased} solvers, 
non-integer numbers were encoded in two different ways: in binary (when bit vector arithmetics is used) 
and also in real (when using rational arithmetic). The fixed-point representation was performed 
by dividing the number to be represented between its integer part $I$, with $m$ bits, 
and its fractional part $F$, with $n$ bits~\cite{Bessa2016}. This approach is represented within 
the tuple $\left<I,F\right>$, which can be encoded both in bit vectors and rational arithmetic 
and is interpreted as $I + F/2^{n}$. Thus, all the represented values must be between the 
maximum and minimum expected values, that is
\begin{equation}
\footnotesize
 V_{max} = 2^{m} - 1/2^{n},
\end{equation}
\begin{equation}
\footnotesize
 V_{min} = -2^{m}
\end{equation}

\noindent and

\begin{equation}
\footnotesize
 V_{min} \le v_{fixed} \le V_{max}.
\end{equation}

All experiments were conducted with a Intel Core i7-2600 PC, with $3.40GHz$ 
of clock speed and $16GB$ RAM and 64-bits Ubuntu as operational system. 
The verification times presented in the following tables are related to 
the average CPU time measured with the times system call (POSIX system) 
of 20 consecutive executions for each benchmark, where the measurement 
unit is always in seconds.

%--------------------------------------
\subsection{Experimental Objectives} \label{exp2}
%--------------------------------------

While creating our benchmark, we aimed for the variety of filter parameters, such as order, frequency and type, so we could demonstrate the usefulness of the method in all sorts of situations. Some filters were defined with an extremely short interval of passband, so that DSVerifier was taken to the limit when trying to represent unrealistically challenging situations. 

%--------------------------------------
\subsection{Results} \label{exp3}
%--------------------------------------

%This subsection describes the results obtained when the benchmarks are 
%executed with the DSVerifier tool modified to 
%support the ESBMC methodology proposed in this work, 
%for the verification of digital systems. 

Table I summarizes the verification results of phase and magnitude, with the first letter in the identification name indicating which filters are FIR and which are IIR.  Filters that contain ``hp'' are high-pass, while ``lp'' stands for low-pass. The filter order is displayed as the number in each filter identification name. Columns ``CF'', ``PF'' and ``SF'' indicate, respectively, the cut-frequency, the pass-frequency and the stop-frequency in kHz used when synthesizing the filters. Some of these frequencies are not employed (indicated by ``NE'') in the design specifications of certain filter types. ``VTM'' stands for the verification time for magnitude, while ``SM'' stands for the magnitude verification status. The status can be either Successful (S), Passband fail (FP), Stopband fail (FS) or Cutoff-frequency fail (FC) . SP is the status for phase verification. The phase status can be either Successful (S) or Fail (F). FP represents the quantity of bits used for fixed-point representation. The minimum gain, for the passband, is fixed in $-1dB$ in Eliptic filters, while the maximum gain for stopband is fixed in $-80dB$ for Eliptical and Chebychev filters. These specifications apply for both low-pass and high-pass filters. Near identical criteria were used for band-pass filters, except that now a pair of each region frequencies is needed too the fully specification of the project. All filters in table I were verified considering a fixed-representation tuple of$\left <4,10\right>$, except by the second order filters, where  $\left<1,5\right>$ was used.

The results for magnitude verification of band-pass filters are included on table II. It's important noting that instead of a single frequency as specification, now a frequency pair is used. The column ``FP tuple'' indicates the considered FWL constriction. 

The poles and zeros verification occurred in the set of IIR filters and the result is shown in table III. ``VT'' stands for the verification time, in seconds, and ``SPZ'' represents the status of the verification. It is worth noting the results for filters with cutoff-frequency of $100Hz$ since magnitude failures were found for all IIR filters.
Table IV presents the results of overflow verifications, which indicates the efficacy on the detection of this type of error. All the tested filters were IIR, that were also used in Table I.

\begin{table}
\scalefont{0.8}
\centering
    \caption{Magnitude and Phase Verification of IIR and FIR filters.}
    \begin{center} {
    \begin{tabular}{|l|l|l|l|l|l|l|l|l|l|l|l|l|l|l|l|l|l|l|}
    \hline

  IIR filters & CF& PF & SF & TVM & SM  & SP \\ \hline
  ilp2 & 9.6 & Na & Na & 5.25 & S  & S  \\ \hline %1
  ihp2 & 9.6 & Na & Na & 5.39 & S & S  \\ \hline %2 
  ilp2EST & 0.1 & Na & Na & 207.34 & FP & F  \\ \hline %3 NAN
  ilp12 & 9.6 & Na & Na & 17.07 & S  & F \\ \hline %4
  
  ihp12 & 9.6 & Na & Na & 17.15 & S  & F  \\ \hline %5
  ilp12EST  & 0.1 & Na & Na & 423.56  & FP & F  \\ \hline %6 NAN
  ilp4C & Na & 9.6 & Na & 173.44 &  FP  & F  \\ \hline %7
  ihp4C & Na & 9.6 & Na & 162.24 & FS  & S  \\ \hline %8
  ilp4ESTC& Na & 0.1 & Na & 163.70 & FS  & F  \\ \hline %9

  ilp12C  & Na & 9.6 & Na & 430.85 & FS  & F  \\ \hline %10
  ihp12C  & Na & 9.6 & Na & 432.66 & FS  & S  \\ \hline %11
  
  ilp12ESTC  & Na & 0.1 & Na & 523.98 & FP  & F  \\ \hline %12 NAN
  ilp4E & Na& Na & 9.6 & 161.44 & FP  & S \\ \hline %13
  ihp4E  & Na & Na & 9.6 & 164.64 & FP & F  \\ \hline %14
  ilp4ESTE & Na & Na & 0.1 & 164.68 & FS  & F \\ \hline %15
  ilp12E  & Na & Na & 9.6 & 460.18 & FP  & F  \\ \hline %16
  ihp12E & Na & Na & 9.6 & 429.14 & FP  & F  \\ \hline %17
  ilp12ESTE & Na & Na & 0.1 & 455.35 & FP  & F  \\ \hline %18

  fhp10 & 7.2 & Na & 9.6 & 14.63 & S  & S  \\ \hline
  flp10 & 9.6 & Na & 9.6 & 1902.8 & FC  & S  \\ \hline
  fhp30 & 9.6 & Na & 9.6 & 1164.9& S  & S  \\ \hline
  flp30 & Na & Na & 9.6 & 41.02 & S  & S  \\ \hline
  flp10EST & 0.1 & Na & 0.1 & 14.95 & S & F  \\ \hline
  flp30EST & Na & Na & 0.1 & 41.91 & S  & S  \\ \hline
 
  fhp10Equi & 9.6 & Na & 9.6 & 360.9 & FS  & S  \\ \hline
  flp10Equi & 9.6 & Na & 9.6 & 70.8 & FP & S  \\ \hline
  fhp30Equi & 9.6 & Na & 9.6 & 1043.2 & FP  & S  \\ \hline
  flp30Equi & Na & Na & 9.6 & 1061.3 & FS  & F  \\ \hline
  flp10ESTEqui & 0.1 & Na & 0.1 & 363.59 & FP & F  \\ \hline
  flp30ESTEqui & Na & Na & 0.1 & 1052.6 & FP  & S  \\ \hline

  fhp10Hann & 9.6 & Na & 9.6 & 361.84 & FC & F \\ \hline
  flp10Hann & Na & Na & 9.6 & 15.42 & S & S \\ \hline
  fhp30Hann & 9.6 & Na & 9.6 & 41.91 & S & S  \\ \hline
  flp30Hann & 9.6 & Na & 9.6 & 40.25 & S & F  \\ \hline  
  flp10ESTHann & Na & Na & 0.1 & 15.36 & S & F  \\ \hline
  flp30ESTHann & Na & Na & 0.1 & 41.22 & S & S \\ \hline

    \end{tabular} } \vspace*{-4ex}
    \end{center}
    \label{table:resultados-mag-fase-IIR}
\end{table}

\begin{table}
\scalefont{0.8}
\centering
    \caption{Magnitude Verification of IIR Passband Filters.}
    \begin{center} {
    \begin{tabular}{|l|l|l|l|l|l|l|l|l|l|l|l|l|l|l|l|l|l|l|}
    \hline

  IIR filters & CF Pairs& PF Pairs & SF Pairs & TVM & SM & FP tuple \\ \hline
  ipb2 & 7.2, 16.8 & Na & Na & 106.46 & FP  & \textless1,5\textgreater \\ \hline
  ipb12 & 7.2, 16.8 & Na & Na & 17.20 & S & \textless4,10\textgreater \\ \hline
  ipb12EST & 7.2, 7.32 & Na & Na & 349.95 & FP  & \textless4,10\textgreater \\ \hline

  ipb4E & Na & 7.2, 16.8 & Na & 160.24 & FP  & \textless4,10\textgreater \\ \hline 
  ipb12E & Na & 7.2, 16.8 & Na & 447.75 & FP  & \textless4,10\textgreater \\ \hline
  ipb12ESTE & Na & 7.2, 7.32 & Na & 438.95 & FS  & \textless4,10\textgreater \\ \hline
 
  ipb4C & Na & Na & 7.2, 16.8 & 140.27 & FP  & \textless4,10\textgreater \\ \hline
  ipb12C & Na & Na & 7.2, 16.8 & 564.95 & FS  & \textless10,16\textgreater \\ \hline
  ipb12ESTC & Na & Na & 7.2, 7.92 & 445.73 & FS  & \textless10,16\textgreater \\ \hline

    \end{tabular} } \vspace*{-4ex}
    \end{center}
    \label{table:resultados-mag-fase-IIR}
\end{table}

\begin{table}
\scalefont{0.8}
    \caption{Stability verification of IIR filters.}
    \begin{center} {
    \begin{tabular}{|l|l|l|l|l|l|}
    \hline
  IIR filters & O & FC (Hz) & Time (s) & SPZ & FP tuple \\ \hline
  hp12 & 12 & 9600 & 1.35 & S & \textless4,10\textgreater \\ \hline
  hp12C & 12 & 9600 & 1.35 & S & \textless4,10\textgreater \\ \hline
  hp12E & 12 & 9600 & 79.99 & F & \textless4,10\textgreater \\ \hline
  hp2 & 2 & 9600 & 1.20 & S & \textless1,5\textgreater \\ \hline
  hp4E & 4 & 9600 & 1.52 & S & \textless4,10\textgreater \\ \hline
  lp12 & 12 & 9600 & 1.35 & S & \textless4,10\textgreater \\ \hline
  lp12C & 12 & 9600 & 1.30 & S & \textless4,10\textgreater \\ \hline
  lp12E & 12 & 9600 & 79.83 & F & \textless4,10\textgreater \\ \hline
  lp12ESTC & 12 & 100 & 1.88 & F & \textless4,10\textgreater \\ \hline
  lp12ESTE & 12 & 100 & 79.94 & F & \textless4,10\textgreater \\ \hline
  lp2 & 2 & 9600 & 1.16 & S & \textless1,5\textgreater \\ \hline
  lp4E & 4 & 9600 & 1.22 & S & \textless4,10\textgreater \\ \hline
    \end{tabular} } \vspace*{-4ex}
    \end{center}
    \label{table:resultados-polos-zeros}
\end{table}

\begin{table}
\scalefont{0.8}
    \caption{Overflow verification of IIR filters.}
    \begin{center} {
    \begin{tabular}{|l|l|l|l|l|}
    \hline
  Filter & Order & Time (s) & FP tuple & Status \\ \hline
  lp2    & 2  & 77.184   & \textless1,5\textgreater & S \\ \hline
  hp2    & 2  & 90.034   & \textless1,5\textgreater & F \\ \hline
  lp4E   & 4  & 185.945 & \textless1,5\textgreater & F \\ \hline
  hp4E   & 4  & 184.254   & \textless1,5\textgreater & F \\ \hline
    \end{tabular} } \vspace*{-4ex}
    \end{center}
    \label{table:resultados-overflow} \vspace*{-2ex}
\end{table}

\section{Conclusion}
The present work proposed a methodology for the verification of digital filter design parameters through the BMC technique, which indicates if the amount of bits used in the representation of the coefficients and samples changes the previously specified characteristics. During the simulations, both IIR and FIR filters were addressed, in order to ensure the outcome of real projects, in different realizations and applications. The results show that it is possible to detect low and medium order filters with a moderate verification time. 
The main contributions of this work are the incorporations of parameters related to filter design theories such as magnitude and phase responses and the location of poles and zeros, which complements the more computational tests, such as overflow, as well as adding support for the filter related verifications in the DSVerifier tool. For future work, it would be of interest incorporating other parameters and the automatic acquisition of the minimum number of bits for the successful validation of the filter design, at fixed-point.

\section*{Aknowledgments}
The authors would like to thank FAPEAM, CNPq, IRCICA and CAPES, for the financial aid.




%\bibliographystyle{IEEEtran}

% \end{thebibliography}

\end{document}